# Formation and Thermal Stability of sub-10 nm Carbon Templates on Si(100)

Olivier Guise[1,3,4], Joachim Ahner[3,4,5], John T. Yates, Jr[1, 2,3,4], and Jeremy Levy[2,3,4,+]

Department of Chemistry[1] Department of Physics and Astronomy[2]

Surface Science Center[3]

Center for Oxide Semiconductor Materials for Quantum Computation[4]

University of Pittsburgh, Pittsburgh, PA 15260

Tel: 412-624-8320, FAX: 412-624-6003

[5]Seagate Technology, Pittsburgh, PA 15222

[+] jlevy@pitt.edu

Abstract:

We report a lithographic process for creating high-resolution (<10 nm) carbon templates on Si(100). A scanning electron microscope, operating under low vacuum ($10^{-6}$ mbar), produces a carbon-containing deposit ("contamination resist") on the silicon surface *via* electron-stimulated dissociation of ambient hydrocarbons, water and other adsorbed molecules. Subsequent annealing at temperatures up to 1320 K in ultra-high vacuum removes $SiO_2$ and other contaminants, with no observable change in dot shape. The annealed structures are compatible with subsequent growth of semiconductors and complex oxides. Carbon dots with diameter as low as 3.5 nm are obtained with a 200 µs electron-beam exposure time.

Control over the growth of silicon-based structures at the sub-10 nm scale approaches the end of scaling for silicon. Single electron devices[1-6], quantum cellular automata[7,8], and quantum computing architectures[8-10] require device dimensions that fall below the 10-nm limit. Atomic-scale assembly in Si has already been demonstrated using scanning tunneling microscopy[11,12] (STM). However, STM-based self-assembly is inherently both demanding and time consuming. Self-assembled quantum dots, produced by Stranski-Krastanow growth processes[13,14], exhibit control over the size but not the location of islands.

For decades it has been recognized that background gases in a scanning electron microscope (SEM) can be used as a "contamination resist" to create structures as small as 8 nm[15,16]. The lateral size of the deposited structures is determined by the convolution of the beam size with the spatial range of secondary electrons.

We have developed a procedure for creating carbon templates on Si(100) that will be employed for the subsequent growth of semiconductors or complex oxides, with reproducible feature sizes below 10 nm. These techniques may prove useful in the growth of controlled arrays of semiconductor quantum dots, devices that use few or single electrons, and quantum computing architectures[9]. The techniques described here are compatible with subsequent UHV-growth of materials using molecular beam epitaxy (MBE) or other growth methods.

For the experiments described here, the following cleaning procedure is used. Si(100) crystals (boron doped, $\rho$=0.05 $\Omega$·cm, single-side polished) are cut to dimensions 10x5x0.5 mm$^3$. Figure 1 shows schematically the sample preparation procedure. The polished Si surface is ozone-cleaned in air for 20 min. using a low-pressure mercury lamp, followed by a modified RCA *ex-situ* chemical cleaning procedure[17]: (1) 10 min. ($H_2O_2$:$H_2SO_4$ 1:2) at 130°C [18]; (2) Standard Clean-1 at 65°C for 5 min. ($H_2O$:$H_2O_2$:$NH_4OH$ 5:1:1)[19]; (3) Standard Clean-2 at 65°C for 10 min. ($H_2O$:$H_2O_2$:HCl 6:1:1) [19,20]; (4) oxide etch for 15 seconds using (HF:$H_2O$ 2:100). This treatment is known to produce a hydrogen-terminated Si(100) surface free of organic contaminants[17]. Throughout the process we use water with a resistivity of 18.1 MΩ.cm which is produced by a four-stage deionization process combined with UV-oxidation and a 0.2 micron filter producing low organic contamination (<1 ppb).

Carbon patterns are created using a high-resolution scanning electron microscope (HR-SEM, LEO VT1530), combined with a high-throughput 50MHz pattern generator/ lithography system (Xenos semiconductor) and a ultra-fast beam blanker (Raith). The HR-SEM has a beam diameter of ~1 nm and is operated at an acceleration voltage of 20 kV and a beam current of 340 pA. The background pressure of $10^{-6}$ mbar in the HR-SEM is produced by a turbomolecular pump and backed by an oil-free roughing pump. Carbon is deposited locally on the substrate by electron-beam induced deposition[21] using background gas containing CO and hydrocarbon molecules adsorbed on the Si(100) surface as a precursor[22]. These carbon-containing molecules are decomposed by an electronic excitation process and form amorphous carbon-containing deposits – likely containing O and H - on the surface[23]. Electron-beam-induced deposition carbon-dot formation was investigated by HR-SEM and Atomic Force Microscopy (AFM).

Carbon patterns are created on the surface by controlling precisely the e-beam location and exposure time. The electron fluence can be as small as $4.3 \times 10^{16}$ e.cm$^{-2}$ (430 e.nm$^{-2}$) which corresponds to an exposure time of 160 ns, although typical exposure times are in the 100 μs -10 ms range. Larger-scale patterns are created by stepping the coarse position of the sample stage.

Figure 2 shows examples of carbon dot patterns formed by the procedure described above. Figure 2(a) features an HR-SEM image of a square pattern of carbon dots. It shows the regularity of the interdot spacing, l=50 nm, with a positional accuracy of ±1 nm in the 100-1000 nm range. Figure 2(b) shows a tapping-mode AFM scan of a different sample with l=35 nm. The dots themselves exhibit a high degree of uniformity in both size and shape, as determined by both HR-SEM and AFM. The exposure time is 1000 μs for Figure 2(a) and 1000 μs for Figure 2(b).

While AFM overestimates the width of the carbon dots due to the AFM probe interaction with the features, the height of the dots can be extracted from the AFM images. Figure 2(d) shows atomic force microscopy cross-sections of the high-density array of carbon dots produced with 1000 μs exposure time (dot-to-dot spacing ~35 nm) imaged in Figure 2(b), revealing their height (~2.1 nm average). On the other hand, height information cannot be obtained from SEM images, but an upper bound of the width may be derived from analysis of the SEM images taken after patterning. Figure 2(c) shows a FWHM ~9.1 nm (estimated from the SEM image) for a pattern produced with 1000 μs exposure time.

Analysis of the dot height shows that the amount of carbon deposited is directly related to the beam exposure time (Figure 3). Carbon dots with various exposure times have been patterned on one single wafer. Over most of the range explored (200 μs to 1 s), the height $h$ scales with the exposure time $t$ according to $h \propto t^{1/3}$. The dependence of the dot width $w$ (measured by AFM) agrees well with $w \propto t^{1/3}$ only at large exposure times.

The deviation is most likely due to imaging artifacts associated with the finite radius of the AFM probe. Assuming that convolution of the dot profile with the tip simply adds a fixed amount to the measured diameter, one can fit the observed results to the $t^{1/3}$ scaling law, also shown in Figure 3. Under the conditions described above, the growth of the carbon dots appears isotropic, with a minimum dot diameter of 3.5 nm and 1 nm dot height obtained with a 200 µs exposure time.

For subsequent processing it is imperative that the carbon dots be stable to high-temperature annealing. In particular, it must be possible to remove oxide layers in order to grow subsequent semiconducting or oxide structures in UHV. In our annealing experiments we first treat the dot pattern with ozone to remove hydrocarbon contaminants from the substrate. Ozone cleaning prevents surface roughening by creating a protective oxide layer on the substrate[24]. Figure 4 shows AFM images of carbon dots before and after UHV-annealing at a pressure of 5.0 x 10$^{-10}$ mbar and temperature of 1220 K for 5 minutes. The carbon pattern retains its overall geometry –symmetry and spacing. The gross pattern appearance did not change up to 1320 K, above which the dots disappeared, presumably due to migration of the carbon features into the bulk. We also observed a slight reduction of the dot height by AFM after ozone-cleaning of the dots (not shown).

Electron-stimulated carbon deposition on Si(100) has been investigated using various pure hydrocarbon gases and was studied in UHV by AES, XPS and TPD[25]. The thermal stability of the carbon films and dots was studied over the temperature range from 300 K – 1400 K, where it was shown that the amorphous carbon begins to convert to SiC at 989 K. This SiC structure is consistent with the observed high thermal stability of the carbon patterns.

In conclusion, sub-10 nm carbon patterns have been created lithographically on Si(100) by electron-beam induced deposition. These carbon dots are stable at temperatures higher than that required for the evaporation of SiO from oxidized silicon surfaces, making them very suitable for subsequent growth of semiconductors and oxides. We expect to employ locally strained Si surfaces[26] for subsequent semiconductor quantum dot growth [27,28] in organized patterns by directed self-assembly as well as for producing nanostructured ferroelectric materials.

The authors gratefully acknowledge the financial support of DARPA QuIST through ARO contract number DAAD-19-01-1-0650, a DARPA grant (DAAD16-99-C-1036), the Department of Chemistry at the University of Pittsburgh for AFM facilities and a DURIP grant from ARO.

**Figure Captions**

Figure 1: Process for carbon patterning of a Si(100) wafer. Wafer is (a) ozone-cleaned using a low pressure mercury lamp; (b) H-passivated using an RCA-derived chemical cleaning technique; and (c) transferred to a high-resolution SEM where ultra-small carbon dots are patterned on the Si surface.

Figure 2: (a) SEM Image of a 5 x 5 array of ultra-small carbon dots. Spacing between the dots is 50nm. (b) Contact-mode AFM image of a similar carbon-dot array. Spacing between the dots is 35nm. (c) FWHM profile of carbon dots from SEM image of Figure 2.a. Average FWHM is ~ 9 nm. (d) Cross section profile of AFM from Figure 2.b. showing an average height ~2.1 nm for the carbon dots. The dots were deposited by 1000 µs irradiation.

Figure 3: Height (squares) and FWHM (circles) of carbon dots, measured by AFM, versus electron beam exposure time. The FWHM by AFM was corrected (stars) assuming a tip radius of 5nm.

Figure 4: AFM images of a carbon dot array before (left) and after (right) annealing in ultra-high vacuum (base pressure 5 x $10^{-10}$ mbar) at 1220K. The carbon dot spacing on appearance is stable up to 1320K.

3-Step Process for High-resolution Carbon Dot Patterning on Si(100)

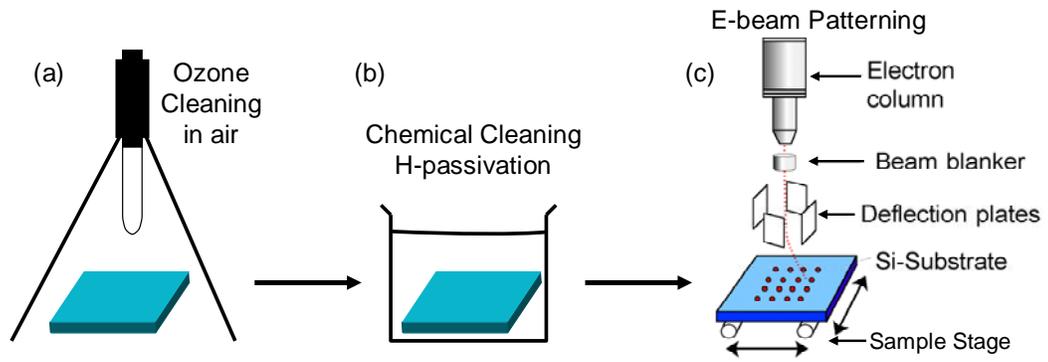

Figure 1

SEM and AFM Imaging of Carbon Dot Patterns

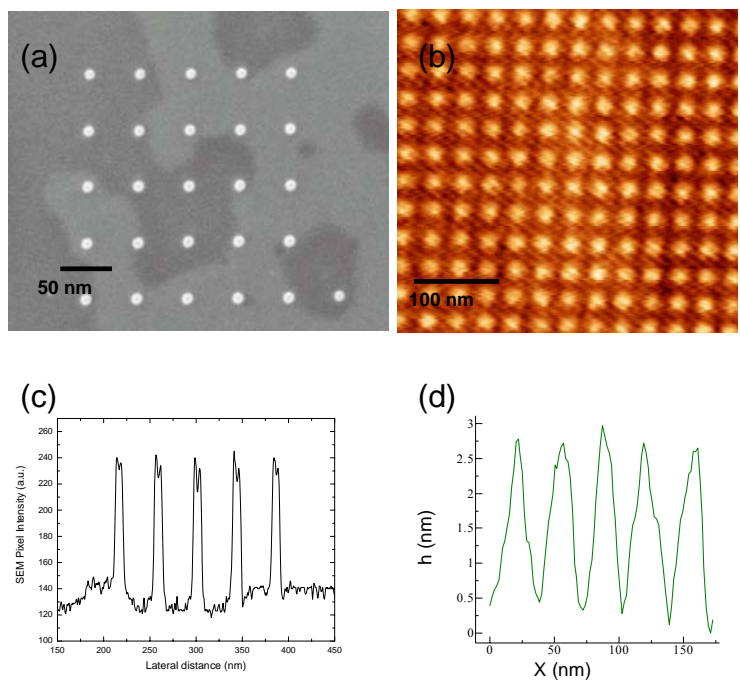

Figure 2

Variation of Carbon Dots Height and Width (obtained
from AFM Images) *vs.* Exposure Time

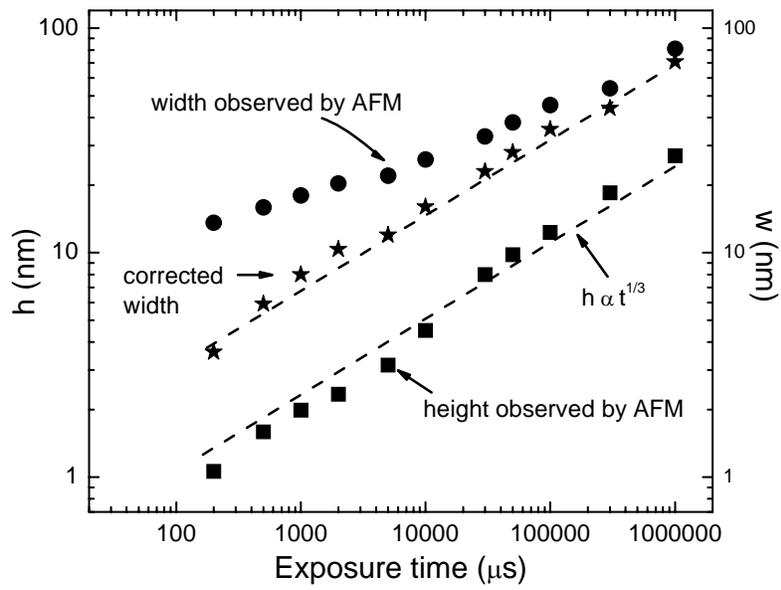

Figure 3

Invariance of Carbon Pattern Upon UHV-Annealing

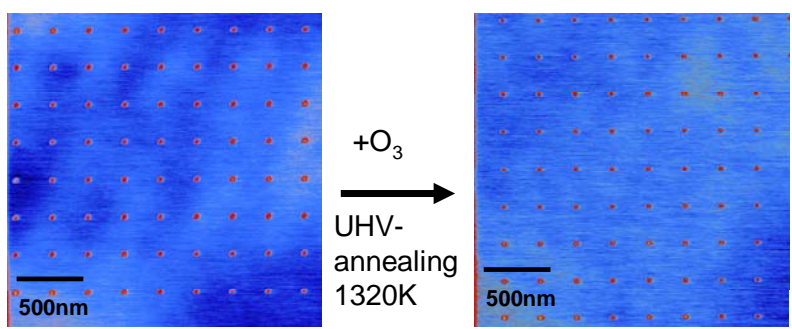

+O₃

UHV-
annealing
1320K

Figure 4